\documentclass[aps,preprintnumbers, amsmath, showpacs]{revtex4}
\usepackage{psfrag}
\usepackage{epsfig}
\usepackage{amsfonts} 
\usepackage{amssymb}  
\usepackage{graphicx} 
\usepackage{amsmath}  
\usepackage{amsmath}  
\usepackage{slashed}
\usepackage[latin1]{inputenc}
\usepackage{color}
\usepackage{suffix}
\usepackage{mathtools}

\begin{document}
\title{No Net Charge Separation in Hot QCD in a Magnetic Field}
\author{E. J. Ferrer and V de la Incera}
\affiliation{Dept. of Physics and Astronomy, CUNY-College of Staten Island and CUNY-Graduate Center, New York 10314, USA}

\begin{abstract}
We study the realization of axion electrodynamics in QCD in the presence of a background magnetic field at temperatures high enough for the occurrence of topological charge transitions that are reflected in the presence of a $\theta$-vacuum term in the action. We show that in this system, the Maxwell equations contain two equal and opposite electric currents that are proportional to the time derivative of the axion field $\theta$. One of these currents comes directly from the Abelian chiral anomaly term in the action and can be interpreted as a polarization current due to the magnetoelectricity of the system with CP-broken symmetry. The other current is obtained from the regular tadpole diagrams and can be understood as produced by the medium chiral imbalance and the single spin projection of the quarks in the lowest Landau level.  Since the two currents cancel out, the net electric charge separation along the magnetic field, a phenomenon known as the Chiral Magnetic Effect, does not take place in hight-T QCD at least in equilibrium, in sharp contrast with many claims in the literature. We discuss the similarities and differences with Weyl semimetals in a magnetic field.\end{abstract}

\pacs{12.38.-t, 12.38.Ki, 05.30.Mh, 71.27.+a}


\maketitle

\section{Introduction}

In recent years, the discovery of anomalous transport in theoretical studies of QCD under extreme conditions and in the presence of electric and magnetic fields has opened the possibility to connect microscopic properties of new quark matter phases to observable macroscopic effects, thus becoming a hot topic of investigation \cite{Huang1509}. The anomalous effects have included the Chiral Magnetic Effect (CME) \cite{CME1}-\cite{Kharzeev-75-14}; the Chiral Separation Effect (CSE) \cite{CSE};  the Chiral Electric Separation Effect (CESE) \cite{CESE}; and most recently, the Anomalous Hall Effect (AHE) \cite{Ferrer-Incera-PLB}-\cite{Universe}. The CME has been predicted to occur in the Quark-Gluon Plasma (QGP) in a magnetic field when there is chiral imbalance, which manifests as a nonzero chiral "chemical potential" $\mu_5$. The CSE is also predicted to occur in hot QGP in a magnetic field, but at finite density, thus with a nonzero baryon chemical potential $\mu$. The CESE requires $\mu_5$, $\mu$, and a background electric field. The AHE is predicted to exist in a phase of cold quark matter at finite density in the presence of a magnetic field, so it requires a nonzero $\mu$.  Interestingly enough, some of these anomalous transport phenomena have similar counterparts in a very different context: topological materials as topological insulators (TI) and Weyl semimetals (WSM) \cite{Rev}.

In this paper we reconsider the phenomena of CME. We are going to work with a chirally unbalanced QGP in the presence of a background magnetic field. Let us recall the arguments on how a chiral imbalance can be generated in the hot QGP. As known, the QCD vacuum is made of an infinite number of topologically inequivalent gluon-vacuum configurations, each characterized by a topological charge or winding number, and separated from each other by an energy barrier. Different vacua can be connected via quantum tunneling by Euclidean gauge-field configurations (instantons) that go to vacuum solutions of different topological charge at $\pm$ infinity, thereby inducing interesting P-odd effects \cite{Belavin}-\cite{ t-Hooft}. At finite temperature, instantons are suppressed by color screening and play no role.  However, in the hot QGP, a parity-odd environment can be generated via sphaleron transitions over the energy barrier that separates the topologically inequivalent vacua \cite{sphaleron}. 

The way to account for these transitions in the action is by adding to the QCD Lagrangian a theta vacuum or axion term that takes the form of a non-Abelian axial anomaly term $\theta G^a_{\mu\nu}\tilde{G}_a^{\mu\nu}$, with $\theta$  a pseudoscalar function known as the $\theta$-angle or the axion field. Here, $G^a_{\mu\nu}$ is the gluon field tensor and $\tilde{G}_a^{\mu\nu}$ its dual. For massless quarks, a constant $\theta$ gives no observable consequence because the axion term can be rotated away by a $U_A(1)$  transformation, but if $\theta$ is spacetime dependent, it can lead to important physical consequences. 

A time-dependent axion field gives rise to an asymmetry in the number of right- and left-handed quarks via the chiral anomaly.  In the CME, a $\frac{\theta}{2N_f}=\mu_5t$ with constant $\mu_5$ has been considered \cite{CME}. Despite $\mu_5$ has been termed "the chiral chemical potential," this terminology is not, strictly speaking, quite accurate. The reason is that in the grand canonical approach, a chemical potential is by definition the thermodynamic conjugate of a conserved charge. However, $\mu_5$ cannot be the conjugate to the axial charge $\bar{\psi}\gamma^0\gamma_5 \psi$ simply because this charge is anomalous, thus not conserved at the quantum level.

The hallmark of the CME can be summarized then by the statement that if $\frac{\theta}{2N_f}=\mu_5t$ exists in the presence of a background magnetic field, the generated chiral imbalance, combined with the single spin projection of the massless quarks in the Lowest Landau level (LLL), lead to a net electric current parallel to the magnetic field direction and thus to a net separation of the electric charges along the direction of the magnetic field. 

The topological origin of the CME hints that the effect should last even at strong coupling. From a theoretical point of view, the limit of strong coupling has been argued to be accessible through the holographic correspondence, what motivated investigating the CME in this context \cite{holographic} - \cite{Rubakov}. Interestingly, the CME  was found to vanish in a gauge theory with a gauge invariant ultraviolet regularization \cite{Holographic-No-CME}.  Later on, it was argued that the net CME current induced by  a constant background axial vector $A^A_0$ and a magnetic field is indeed zero, but if instead, one introduces a chiral chemical potential for the \textit{conserved axial charge}, meaning the one that incorporates the regular and anomalous terms together, then the CME current is regained \cite{Rubakov}.  This last point is nevertheless questionable. As known, to construct the many-particle theory, one has to introduce in the Hamiltonian $H$ all the classically conserved charges (obtained from the Noether theorem) multiplied by their corresponding chemical potentials, $H(\mu)=H-\sum_i \mu_i J_i$. $H(\mu)$ is then inserted in the partition function of the quantum-statistic theory and use it to calculate all the physical quantities. What was proposed in \cite{Rubakov} equals to construct the partition function with a Hamiltonian that already depends on a quantum correction given by the anomalous term.

In hot QCD, however, we do not have to deal with this kind of potentially disputable argument, since in this case $\mu_5$ enters in the theory only through the theta vacuum term, which as argued above is a consequence of the nontrivial QCD vacuum and the sphaleron-induced transitions between topologically inequivalent vacuum configuration. It is in this framework where we are going to revisit the CME phenomenon with the goal to shed new light on the debate about its existence. We will base our analysis on a systematic derivation of all the possible contributions to the electric current. With that aim in mind, we perform a $U^A(1)$ local axial transformation of the fermion fields that allows to transfer the axial anomaly from the non-Abelian (gluonic) sector to the Abelian (electromagnetic) sector of the theory. In other words, it eliminates the original term $\sim \theta G^a_{\mu\nu}\tilde{G}_a^{\mu\nu}$ and produces a new Abelian axion term $\sim \theta F_{\mu\nu}\widetilde{F}^{\mu\nu}$ in function of the electromagnetic field $F_{\mu\nu}$ and its dual. To show this, we use the Fujikawa's approach \cite{FujikawaPRD21_1980, Fujikawa_book} to regularize the fermion Jacobian produced by the lack of invariance of the fermion measure under the local axial transformation. Once this is done, one can integrate out the fermions to find the electromagnetic action of the theory and then derive the Maxwell equations. The electromagnetic charges and currents are thus readily obtained in a way similar to the one used in high-dense QCD  \cite{Ferrer-Incera-PLB, Ferrer-Incera-NPB}.

Following the procedure described above, we find that the Maxwell equations in the hot QGP in a magnetic field are actually those of Axion Electrodynamics. As a consequence, in addition to the ordinary current -found from the tadpole diagram and depending on $\mu_5$ through the modified spectrum of the LLL fermions - an extra, anomalous current coming from the term $\sim \theta F_{\mu\nu}\widetilde{F}^{\mu\nu}$ also appears.  These two currents are equal and opposite, thus cancelling each other out. This implies that there is no net CME current and therefore, there is no net charge separation in the hot QGP in a magnetic field.

We call the readers' attention to the fact that the situation is different for dense QCD in the Magnetic Dual-Chiral-Density-Wave (MDCDW) phase, where the axion field  $\theta( \mathbf{x})$ only depends on the spatial coordinate \cite{KlimenkoPRD82}. In the MDCDW phase, the anomalous current is not cancelled out by the ordinary one. In fact, in this case the anomalous current turns out to be a dissipationless Hall current with important transport consequences \cite{Ferrer-Incera-PLB, Ferrer-Incera-NPB}. 

The paper is organized as follows. In Section II, we introduce the equations of Axion Electrodynamics for a general axion field $\theta$ and highlight the different contributions to the total 4-current, the anomalous, coming from the axion term of the action, and the ordinary, obtained from tadpole and polarization operators diagram. In Section III, we define the QCD$\times$QED Lagrangian density with the QCD $\theta$-vacuum contribution and discuss its topological characteristics and how it is related to the axial anomaly. In Section IV, we study in detail the Fujikawa approach for the partition function under a local chiral transformation of the QCD$\times$QED model. Then, in Section V, we calculate the corresponding charge and current densities (anomalous and ordinary) for this theory in the presence of a magnetic field, and show that these two currents are equal in magnitude but opposite in direction. In Section VI, we briefly discuss the phenomena of CME and AHE in Weyl semimetals, and highlight the analogies and differences with  topological QCD systems. Finally, Section VII states our concluding remarks.
  
\section{Axion Electrodynamics}

Recently, new macroscopically observable quantum effects that manifest through the interaction of matter with electromagnetic fields in QCD and condensed matter physics have become a focus of attention \cite{Huang1509}-\cite{Rev}. These effects are connected to the nontrivial topology of these systems and are related to parity and/or time-reversal symmetry violation. The interaction between the electromagnetic field and matter with nontrivial topology is described by the equations of axion electrodynamics initially  proposed by Wilzcek \cite{axionElect} to describe the effects of adding a general axion term $\frac{\kappa}{4}\theta F_{\mu\nu}\widetilde{F}^{\mu\nu}$ to the ordinary Maxwell Lagrangian, 
\begin{eqnarray}
\mathbf{\nabla} \cdot \mathbf{E}&=&J_0+J^{anom}_0, \label{axQED_1}
 \\
\nabla \times \mathbf{B}-\frac{\partial \mathbf{E}}{\partial t}&=&\mathbf{J}_V+ \mathbf{J}^{anom},  \label{axQED_2}
\\
\mathbf{\nabla} \cdot \mathbf{B}&=&0, \quad \nabla \times \mathbf{E}+\frac{\partial \mathbf{B}}{\partial t}=0,  \label{axQED_3}
\end{eqnarray}   
The anomalous charge and current densities in the first two equations are derived from the axion term $\frac{\kappa}{4}\theta F_{\mu\nu}\widetilde{F}^{\mu\nu}$, and they depend on the axion field $\theta$ as,
 \begin{equation}\label{Anom-Charge}
J^{anom}_0=\kappa \nabla\theta\cdot \mathbf{B}
\end{equation}
 
 \begin{equation}\label{Anom-Current}
 \mathbf{J}^{anom}=-\kappa\Bigg(\frac{\partial \theta}{dt}\mathbf{B}+\nabla\theta\times \mathbf{E}\Bigg ),
\end{equation} 
where the coefficient, $\kappa$, and the axion field, $\theta(\mathbf{x}, t)$, are model-dependent parameters. In  Eqs. (\ref{axQED_1})-(\ref{axQED_2}),  $J_0$ and $\mathbf{J}_V$ are respectively the ordinary charge and current densities found from tadpole and polarization operator diagrams.

 
 In condensed matter, terms of this form have been shown to emerge in: 1) topological insulators (TI) \cite{Qie-PRB78}, where $\theta$ depends on the band structure of the insulator; 2) Weyl semimetals \cite{Weylsemi}, materials with points of degeneracy between two bands, called Weyl nodes, that lie on the Fermi surface. In this case the angle $\theta$ is related to the energy and/or momentum separation between the two nodes; and  3) Dirac semimetals (DM) \cite{Diracsemimet}, with band-touching nodes like the WSM, but the nodes lie out of the Fermi surface.
 
For quark matter, an electromagnetic axion term can be generated via two separate mechanisms, one at high temperature (T) \cite{Warringa}-\cite{Kharzeev-75-14} and the other at high density \cite{Ferrer-Incera-PLB}-\cite{Universe}. At high T, a nontrivial axion field $\theta$ can arise thanks to the sphaleron transitions over the barrier that separates topologically inequivalent vacua \cite{sphaleron}. Even though $\theta$ originally enters coupled to the gluon field, performing a local axial transformation followed by a proper regularization scheme based on the Fujikawa method \cite{FujikawaPRD21_1980}, the non-Abelian axion term is eliminated (See Section IV). At the same time, a new axion term, $\frac{\kappa}{4}\theta F_{\mu\nu}\widetilde{F}^{\mu\nu}$, in the QED sector of the theory emerges. This Abelian axion term couples $\theta$ to the electromagnetic field $F_{\mu\nu}$ and its dual. We will see below that with a time-varying $\theta$ and in the presence of a background magnetic field, such an induced  term leads to a time-dependent medium polarization. 

The mechanism at high density \cite{Ferrer-Incera-PLB}-\cite{Universe}, takes place in the MDCDW phase of dense quark matter \cite{KlimenkoPRD82}, where $\theta$ is related to the modulation $q$ of the chiral condensate as $\theta=qz/2$. The axion field in this case gives rise to a dissipationless Hall current and an anomalous electric charge \cite{Ferrer-Incera-PLB}-\cite{Universe}.

We call reader's attention to an important fact; the generation of an axion term $\frac{\kappa}{4}\theta F_{\mu\nu}\widetilde{F}^{\mu\nu}$ is a necessary but not sufficient condition to have anomalous electromagnetic transport in the system. The actual conclusion depends on whether the anomalous contributions to the four-current are or not cancelled by the ordinary components that enter in the Maxwell equations (\ref{axQED_1})-(\ref{axQED_2}), 
\begin{equation}\label{Total-Charge}
J^{Total}_0=J_0+J^{anom}_0
\end{equation}

\begin{equation}\label{Total-Current}
 \mathbf{J}^{Total}= \mathbf{J}_V+ \mathbf{J}^{anom}
\end{equation} 
Anomalous transport then requires that the ordinary contributions do not eliminate their anomalous counterparts. 

As will be shown in Section V, in high-T QCD in a magnetic field, $\mathbf{J}_V=- \mathbf{J}^{anom}$ and so the net current is zero, in contrast with what occurs in the high-density case, where the ordinary longitudinal current $\mathbf{J}^{long}_V$ is zero, while the ordinary  $\mathbf{J}^H_V$, and the anomalous $\mathbf{J}^{anom}$ Hall currents are both different from zero, but $\mathbf{J}^H_V$ does not eliminate $\mathbf{J}^{anom}$  \cite{Ferrer-Incera-PLB, Ferrer-Incera-NPB}. 

A possible explanation for the lack of anomalous transport in hot QGP in a magnetic field and its existence in the dense matter case can be argued as follows.  At high-T the non-trivial topology of the system comes from the topologically inequivalent gluon vacuum configurations, but the topology never gets reflected in the quark spectrum, which remains symmetric even after the axial local transformation, and hence continues to be topologically trivial.  That is why the topology of the gluon vacuum does not actually manifest in the electromagnetic transport, which occurs via the fermions. In contrast, in the dense quark matter case, the LLL quarks of the  MDCDW phase have an asymmetric spectrum that is characterized by a topological index \cite{PLB743} and thereby produces anomalous electromagnetic transport \cite{Ferrer-Incera-PLB}-\cite{Universe}.

\section{QCD$\times$QED with $\theta$-vacuum term}

 Let us consider the Lagrangian of massless QCD$\times$QED with the contribution of the P and CP-odd $\theta$-vacuum term \cite{Kharzeev-75-14},  
\begin{eqnarray} \label{L_QCD_QED}
\mathcal{L}_{QCD+QED}=&-&\frac{1}{4}G^a_{\mu\nu}G^{\mu\nu}_a-\frac{1}{4}F_{\mu\nu}F^{\mu\nu}-\frac{g^2}{32 \pi^2}\theta G^a_{\mu \nu}\widetilde{G}_a^{\mu \nu}\nonumber
\\
&+&\bar{\psi}[i\gamma^{\mu}(\partial_\mu-igG^a_\mu\frac{\lambda_a}{2}+iQA_{\mu})]\psi,
\end{eqnarray}
Here, $G^a_\mu$ and $A_\mu$ represent the gluon and photon fields respectively,  $\lambda_a/2$ are the color SU(3) group generators in the fundamental representation, $G_{\mu \nu}^a=\partial_\mu G_\nu^a-\partial_\nu G_\mu^a+gf^{abc}G_\mu ^b G_\nu^c$, where the $f^{abc}$-coefficients with $(a, b, c = 1,...,8)$ are the totally antisymmetric structure constant of SU(3), $\widetilde{G}_a^{\mu \nu}=\frac{1}{2}\epsilon^{\mu\nu\rho\sigma}G^a_{\rho\sigma}$, $\psi^T=(u,d,s)$, $Q=diag(q_u,q_d,q_s)=diag(\frac{2}{3}e,-\frac{1}{3}e,-\frac{1}{3}e)$ is a matrix in flavor space that accounts for the electric charge of the quarks. We assume a general pseudoscalar axion field $\theta(\vec{x},t)$. The electromagnetic potential $A^{\mu}$ is formed by the background $\bar{A}_{\mu}=(0,0,Bx^1,0)$, which corresponds to a constant and uniform magnetic field $B$ in the z direction, plus the fluctuation field $\tilde{A}_\mu$. We use the metric $g_{\mu\nu}=diag(1,- \mathbf{1})$ and Levi-Civita tensor $\epsilon^{0123}=1$. 

The QCD classical vacua have to be pure gauge to have minimal energy. In the temporal gauge $G^a_{0}=0$, they are given by
\begin{equation}\label{vacuum-conf}
G_i(\vec{x})=ig^{-1}U^{-1}(\vec{x})\partial_i U(\vec{x}),
\end{equation}
with $U(\vec{x})\in SU(3)$ and $U(\vec{x}) \to 1$ when $\vec{x} \to \infty$. The vacuum configurations are characterized by a topological number $n_w$ ($n_w \in Z$),
\begin{equation}\label{winding number}
n_w=\frac{1}{24 \pi^2} \int d^3x\epsilon^{ijk}tr(U^{-1}\partial_i U)(U^{-1}\partial_j U)(U^{-1}\partial_k U),
\end{equation}
also known as the winding number. As a consequence, there is an infinite set of topologically different classical vacua classified by the integer $n_w$.  As $n_w$ is a topological quantity, continuous deformations of the gauge fields cannot change it.  Hence, it is not possible to go from one vacuum class to another by a continuous transformation without passing by gauge field configurations that are not vacua. This means that the vacuum classes are separated by a finite energy barrier. The actual QCD vacuum is then a superposition of all the $n_w$ vacua \cite{Callan, t-Hooft, QCD-vacuum}. Such a superposition, known as the $\theta$-vacuum, yields to the axion term $\sim\theta G^a_{\mu \nu}\widetilde{G}_a^{\mu \nu}$, with $\theta$ a pseudoscalar, in the Lagrangian (\ref{L_QCD_QED}) \cite{Rev.Mod.Phys.70}.

Gauge field configurations that go to different topological pure gauge fields at $\pm \infty$ are characterized by a nonzero $Q_w=n_w(-\infty)-n_w(\infty)$ and hence can induce a transition from one topological vacuum to another. At zero temperature, such gauge field configurations are the instantons \cite{Instantons}. Instantons induce quantum tunneling between vacua through the energy barrier, which in this case is of order $\mathcal{O}(\Lambda_{QCD}/\alpha_s)$ with $\Lambda_{QCD}$ the QCD scale and $\alpha_s$ the strong coupling constant. At finite temperature the instantons are color screened so that the tunneling effect is practically suppressed at high temperature \cite{Pisarski-Yaffe}. In this regime, the transition between different vacua is induced by thermal excitations called sphaleron \cite{sphaleron} that connect different vacua over the barrier.

On the other hand, taking into account \cite{Nakahara} that for massless quarks the axial anomaly equates
\begin{equation}\label{anomaly-eq}
\partial_\mu J_A^\mu=\frac{g^2}{16\pi^2} G_{\mu \nu}^a\widetilde{G}^{\mu \nu}_a,
\end{equation}
 with $J_A^\mu=\sum_f \langle  \overline{\psi}_f \gamma^\mu \gamma_5 \psi_f\rangle$ the axial four-current, one can show,
integrating (\ref{anomaly-eq}) in the 4-volume, that 
\begin{equation}\label{Int-anomaly-eq}
(N_R-N_L)=-\frac{g^2N_f}{16\pi^2}\int d^4x G_{\mu \nu}^a\widetilde{G}^{\mu \nu}_a,
\end{equation}
with $N_{R/L}$ being the net number of quarks (or minus for antiquarks) with right- and left-handed chirality respectively. The proportionality to the number of massless flavors $N_f$ accounts for the fact that all the massless flavors equally contribute to the anomaly. This result shows that a topologically nontrivial gauge field configuration can create or annihilate the total chirality of fermions. It could explain how in heavy-ion collisions, on an event-by-event basis, the QGP can become chiral.

In summary, these results link the $\theta$-vacuum term in (\ref{L_QCD_QED}) with the possibility of chirality change in high-T QCD due to the existence of gluon configurations with different winding numbers that can be connected by sphalerons. As known, the triangle anomaly also links the axial current with the electromagnetic field, so a relevant question at this point is whether the $\theta$-vacuum term can induce new terms in the electromagnetic sector of the quantum effective action.  In the next section we use Fujikawa's method \cite{FujikawaPRD21_1980}-\cite{Fujikawa_book} to address this question. Among several available methods to derive the anomaly, Fujikawa's, which is equivalent to the heat kernel proof of the relevant index theorem, has the advantage of most directly revealing the topological nature of the problem. 

\section{Effective action in hot QCD in a Magnetic Field}

Our main goal now is to find the QCD$\times$QED effective iaction after performing a local chiral transformation that eliminates the $\theta$-vacuum contribution from the gluon sector. As we will see, this transformation has two consequences. On the one hand, it modifies the fermion spectrum that now becomes dependent on $\theta$.  On the other hand, it does not leave the fermion measure invariant. The corresponding Jacobian is ill-defined and has to be regularized. Using the Fujikawa's method \cite{FujikawaPRD21_1980} to regularize it, we will show that it leads to new $\theta$-dependent contributions into the action electromagnetic sector.

In our derivations, special attention will be paid to the fact that the term  $\sim\gamma^5\partial_\mu \theta$, induced in the covariant derivative by the local chiral transformation, spoils the Hermiticity of the Euclidean Dirac operator. To deal with such a situation, the Fujikawa approach has to be extended in a similar fashion to what has been done in the presence of a chemical potential \cite{JHEP1112_2011}, or in the MDCDW phase of quark matter \cite{Ferrer-Incera-NPB}.

{\subsection{Regularization of the Fermion Jacobian}

Fujikawa's method allows to obtain the regularized Jacobian of a local $U_A(1)$ chiral transformation that does not leave the fermion measure invariant. In the case under study here, such a transformation takes the form
\begin{eqnarray}\label{quark-transformation}
\psi(x) \rightarrow U_A\psi(x)= e^{i\theta \gamma^5/2N_f}\psi(x)\nonumber
\\
\bar{\psi}(x) \rightarrow \bar{\psi}(x)\gamma_0U^{\dag}_A\gamma_0=\bar{\psi}(x)e^{i\theta \gamma^5/2N_f}
\end{eqnarray}
with $N_f$ the numbers of flavors in the theory.

In general we consider $\theta(\vec{x},t)$. We are not interested in a global $U_A(1)$, since if $\theta$ were constant, the $\theta$-vacuum term in (\ref{L_QCD_QED}) would not have observable consequences because the transformation would yield a total derivative of the Chern-Simons current that would not contribute to the equations of motion. 

Under the local $U_A(1)$, the fermion sector of the Lagrangian (\ref{L_QCD_QED}) acquires a $\theta$-dependent chiral coupling
\begin{equation}\label{quark-Lagrangian}
\mathcal{L}_{\psi}=\bar{\psi}[i\gamma^{\mu}(\partial_\mu-igG_\mu+iQA_{\mu}+\mu \delta_{\mu0}+i\gamma^5\partial_\mu\theta/2N_f)]\psi.
\end{equation}

Moreover, the fermion measure in the path integral
\begin{equation}\label{path-int}
\mathcal{Z}[G_\mu^a, A_\mu, \theta]=\int \mathcal{D}\psi \mathcal{D}\bar{\psi}e^{iS_{\psi}(G_\mu^a, A_\mu, \theta)},
\end{equation} 
with $S_{\psi}(G_\mu^a, A_\mu, \theta)$ the fermion part of the effective action, changes to
\begin{equation}\label{MeasureNI}
\mathcal{D}\bar{\psi}(x)\mathcal{D}\psi(x) \to J_{\bar{\psi}}J_{\psi}\mathcal{D}\bar{\psi}(x)\mathcal{D}\psi(x)
\end{equation} 
with Jacobians $J_{\bar{\psi}}=J_{\psi}=(\textrm{Det}U_A)^{-1}\neq1$. Therefore, the chiral transformation (\ref{quark-transformation}) not only modifies the fermion spectrum, but it also adds an extra term, the Jacobian, to the action. 

To calculate the Jacobian we first perform a Wick rotation to Euclidean space,  $dx_0 \rightarrow -idx_4$, $\partial_0 \rightarrow i\partial_4$, $A_0\rightarrow iA_4$, $\gamma_0\rightarrow i\gamma_4$, $\gamma^5 \equiv i\gamma^0 \gamma^1 \gamma^2 \gamma^3=\gamma_E^5=\gamma^1 \gamma^2 \gamma^3 \gamma^4$. The Euclidean $\gamma_{\mu}$ are all anti-Hermitian; the Euclidean metric becomes $g^E_{\mu\nu}= \mathrm{diag}(-1,-1,-1,-1)$. Since there is no mix between the quark flavors, one can perform, without loss of generality, the analysis for a single flavor and incorporate the contributions of all the flavors in the final result. 

Using $\langle x|\mathcal{O} |y\rangle=\delta^4(x-y) \mathcal{O}(x)$, valid for ultra-local integral kernels, the Jacobian can be written as
\begin{eqnarray}\label{Ill-defDet}
( \textrm{Det}U_A)^{-1}&=&e^{-\mathrm{Tr} \ln U_A}=e^{-\int d^4x \langle x |\mathrm{tr} \ln U_A|x\rangle} \nonumber
\\
&=&e^{-\int d^4x \delta^4(0) \frac{\theta(x)}{2N_f}i\mathrm{tr} (\gamma_5)},
\end{eqnarray}
with $\mathrm{Tr}$ meaning functional+matrix trace and $\mathrm{tr}$ just matrix trace. 

The exponent in (\ref{Ill-defDet}) is ill-defined and needs proper regularization. To regularize it, we follow a gauge-invariant approach introduced by Fujikawa many years ago \cite{FujikawaPRD21_1980, Fujikawa_book} and later extended to finite temperature and  density \cite{JHEP1112_2011,  Finite-Temp}. The essence of the Fujikawa's method is to express the Jacobian in the representation of the eigenfunctions of an Euclidean operator that is gauge-invariant and Hermitian (or antiHermitian).  Such a representation preserves the gauge invariance of the theory and ensures that the eigenfunctions are orthogonal and complete, so they have real (imaginary) eigenvalues. In addition, to ensure unitarity, it is essential that the operator used to define the functional space be chosen so as to diagonalize the fermion action. As discussed in \cite{Fujikawa_book}, this condition is important, since a seemingly unitary transformation based on the eigenspace of a gauge-invariant operator that does not diagonalize the action is actually non-unitary.

Further, to regularize the Jacobian one introduces a damping factor in the form of an arbitrary function of the eigenvalues of this operator with a regulator $M$, in such a way that the contributions from the large momenta are regularized when $M \to \infty$. Below, we use the heat-kernel regularization \cite{Nakahara}, which is based on an exponential damping function.

In most cases, the gauge-invariant operator whose eigenfunctions satisfy all the above requirements is the Dirac operator of the theory $\slashed{D}$ \cite{Bilal}. However, in our case, even at zero density ($\mu=0$), the presence in the covariant derivative  of the chiral term $\sim \gamma_5 \partial_\mu \theta$ spoils the Hermiticity of the Dirac operator in the Euclidean space and the Fujikawa approach has to be extended, as known to happen in the presence of a chemical potential \cite{JHEP1112_2011}, or in the MDCDW phase  \cite{Ferrer-Incera-NPB}.
 
In the present system, the Euclidean Dirac operator is $\slashed{D}(\theta)=\slashed{D}+\slashed{D}^A$, with $\slashed{D}=\gamma_\mu(\partial_\mu-igG^a_\mu\frac{\lambda_a}{2}+iQA_{\mu})$ Hermitian, and $\slashed{D}^A=i\gamma_\mu \gamma^5\partial_{\mu}\theta/2N_f-\mu \gamma_4$ anti-Hermitian. Since $\slashed{D}(\theta)$ is neither Hermitian nor anti-Hermitian, its eigenfunctions cannot be used as a suitable representation in the Fujikawa approach. In this case, we follow instead the method discussed in Refs.  \cite{Ferrer-Incera-NPB, FujikawaPRD21_1980, JHEP1112_2011}.  

Consider the positive-semidefinite Hermitian operators $\slashed{D}^\dag(\theta)\slashed{D}(\theta)$ and $\slashed{D}(\theta)\slashed{D}^\dag(\theta)$ and their respective eigenvalue equations
\begin{equation}\label{DdagDeigenf}
\slashed{D}^\dag(\theta)\slashed{D}(\theta)\phi_n=\lambda_n^2\phi_n \qquad \slashed{D}(\theta)\slashed{D}^\dag(\theta)\widetilde{\phi}_n=\widetilde{\lambda}_n^2\widetilde{\phi}_n, 
\end{equation}
whose eigenfunctions satisfy completeness
\begin{equation}\label{Completeness}
\sum_n \phi^\dag_n(x)\phi_n(y)=\delta(x-y) \qquad \sum_n \widetilde{\phi}^\dag_n(x)\widetilde{\phi}_n(y)=\delta(x-y)
\end{equation}
and orthogonality 
\begin{equation}\label{Orthogonality}
\int d_E^4x{\phi}^\dag_n(x)\phi_m(x)=\delta_{nm}, \qquad \int d_E^4x\widetilde{\phi}^\dag_n(x)\widetilde{\phi}_m(x)=\delta_{nm}.
\end{equation} 
conditions and have real eigenvalues $\lambda_n$, $\widetilde{\lambda}_n$, known as the singular values of $\slashed{D}(\theta)$, $\slashed{D}^\dag(\theta)$ respectively. The $\phi_n(\tau, \vec{x})$ and $\widetilde{\phi}_n(\tau, \vec{x})$ are ordinary c-number functions antiperiodic in $\tau$. In (\ref{Orthogonality}), we introduced the notation $\int d_E^4x\equiv \int_0^\beta d\tau \int d^3x$.

It is easy to verify, as in the case studied in \cite{JHEP1112_2011}, that the operators $\slashed{D}^\dag(\theta)\slashed{D}(\theta)$ and $\slashed{D}(\theta)\slashed{D}^\dag(\theta)$ share all the nonzero eigenvalues. To see this, consider a nonzero $\lambda_n$ and let us act with $\slashed{D}(\theta)$ on the first equation of (\ref{DdagDeigenf})
\begin{equation}\label{Dnuevo}
\slashed{D}(\theta)\slashed{D}^\dag(\theta)\slashed{D}(\theta)\phi_n=\lambda_n^2\slashed{D}(\theta)\phi_n
\end{equation}
This means that $\slashed{D}(\theta)\phi_n$ is an eigenfunction $\widetilde{\phi}_n$ of $\slashed{D}(\theta)\slashed{D}^\dag(\theta)$ with eigenvalue $\lambda_n^2$. Similarly, acting with $\slashed{D}^\dag(\theta)$ on the second equation of (\ref{DdagDeigenf}) we find
\begin{equation}\label{Dnuevo1} 
\slashed{D}^\dag(\theta)\slashed{D}(\theta)\slashed{D}^\dag(\theta)\widetilde{\phi}_n=\widetilde{\lambda}_n^2\slashed{D}^\dag(\theta)\widetilde{\phi}_n, 
\end{equation}
Hence, we also see that $\slashed{D}^\dag(\theta)\widetilde{\phi}_n$ is an eigenfunction of $\slashed{D}^\dag(\theta)\slashed{D}(\theta)$ with eigenvalues $\widetilde{\lambda}_n^2$, so  $\lambda_n^2=\widetilde{\lambda}_n^2$.  Then, we define from now on, for nonzero $\lambda_n$, $\widetilde{\phi}_n=\lambda^{-1}_n \slashed{D}(\theta)\phi_n$. 

Now we can expand the fermion fields in the bases of the Hermitian operators $\slashed{D}^\dag(\theta)\slashed{D}(\theta)$ and $\slashed{D}(\theta)\slashed{D}(\theta)^\dag$ as,
\begin{equation}\label{representation}
\psi(x)=\sum_{n} a_n\phi_n(x), \qquad \bar{\psi}(x)=\sum_{n} \bar{b}_n{\widetilde{\phi}}^\dag_n(x),
\end{equation}
with $a_n$, $b_n$, Grassmann numbers. In the representation of these eigenfunctions, the Jacobian of flavor $f$ in (\ref{MeasureNI}) takes the form
\begin{equation}\label{Jacobian-phi-rep}
J^{(f)}_{\psi}J^{(f)}_{\bar{\psi}}=e^{-\frac{1}{2N_f}\mathrm{tr}\int d_E^4x \theta(x) \sum_{n} [\phi^\dag_n(x) i\gamma_5 \phi_n(x)+\widetilde{\phi}^\dag_n(x) i\gamma_5 \widetilde{\phi}_n(x)].}
\end{equation}
and the fermionic part of the action is diagonalized
\begin{equation} \label{diagonal-action}
S_F=\int d_E ^4x\bar{\psi}\slashed{D}(\theta)\psi=\sum_{n}\lambda_n \bar{b}_n a_n.
\end{equation}

We now apply the standard heat-kernel regularization method \cite{Nakahara}, and introduce damping factors for each term in (\ref{Jacobian-phi-rep}) with a regulator $M$ that will be taken to infinity at the end. The regularized Jacobian then becomes
\begin{equation}\label{Jacobian-phi-rep_Reg}
 J^{(f)}_{\psi}J^{(f)}_{\bar{\psi}}
 =e^{-\frac{1}{2N_f}(\mathcal{I}_R+\mathcal{\widetilde{I}}_R)}
\end{equation}
where 
\begin{eqnarray}\label{Heat-Kernel}
\mathcal{I}_R&=&\lim_{M \rightarrow \infty} \int d_E^4x \theta(x)\mathrm{tr}\sum_n\phi_n^{\dag}(x)  i\gamma_5 e^{-\lambda_n^2/M^2} \phi_n(x)  \nonumber
\\
&=&\lim_{M \rightarrow \infty} \int d_E^4x \theta(x)\mathrm{tr}\sum_n\phi_n^{\dag}(x) i \gamma_5 e^{-\slashed{D}^\dag(\theta)\slashed{D}(\theta)/M^2}\phi_n(x) \nonumber
\\
&\equiv& \lim_{M \rightarrow \infty}\int d_E^4x \theta(x)I,
\end{eqnarray}
and
\begin{eqnarray}\label{Heat-Kernel1}
\mathcal{\widetilde{I}}_R&=&\lim_{M \rightarrow \infty} \int d_E^4x \theta(x)\mathrm{tr}\sum_n\widetilde{\phi}_n^{\dag}(x)  i\gamma_5 e^{-\lambda_n^2/M^2} \widetilde{\phi}_n(x)  \nonumber
\\
&=&\lim_{M \rightarrow \infty} \int d_E^4x \theta(x)\mathrm{tr}\sum_n \widetilde{\phi}_n^{\dag}(x) i \gamma_5 e^{-\slashed{D}(\theta)\slashed{D}^\dag(\theta)/M^2}\widetilde{\phi}_n(x) \nonumber
\\
&\equiv& \lim_{M \rightarrow \infty}\int d_E^4x \theta(x)\widetilde{I},
\end{eqnarray}
with
\begin{eqnarray}\label{Square-Opr}
\slashed{D}^\dag(\theta)\slashed{D}(\theta)=&+&\frac{iq_f}{4} [\gamma_\mu, \gamma_\nu] F_{\mu\nu}+\frac{1}{4} [\gamma_\mu, \gamma_\nu] G_{\mu\nu}-i\frac{\mathrm{sgn}(q_f)}{2N_f}\gamma_5\left[\gamma_\mu, \gamma_4\right]\mu\partial_\mu\theta \nonumber
\\
&+&(\frac{\partial_{\mu}\theta}{2N_f})^2-(D_\mu)^2+i \frac{\mathrm{sgn}(q_f)}{2N_f}[\gamma_\mu, \gamma_\nu]\gamma_5\partial_\mu\theta D_{\nu}+\mu^2
\end{eqnarray}
\begin{eqnarray}\label{Square-Opr1}
\slashed{D}(\theta)\slashed{D}^\dag(\theta)=&+&\frac{iq_f}{4} [\gamma_\mu, \gamma_\nu] F_{\mu\nu}+\frac{1}{4} [\gamma_\mu, \gamma_\nu] G_{\mu\nu}-i\frac{\mathrm{sgn}(q_f)}{2N_f}\gamma_5\left[\gamma_\mu, \gamma_4\right]\mu\partial_\mu\theta \nonumber
\\
&+&(\frac{\partial_{\mu}\theta}{2N_f})^2-(D_\mu)^2-i \frac{\mathrm{sgn}(q_f)}{2N_f}[\gamma_\mu, \gamma_\nu]\gamma_5\partial_\mu\theta D_{\nu}+\mu^2
\end{eqnarray}
Here, ${D}_\mu=\partial_\mu-igG^a_\mu\frac{\lambda_a}{2}+iq_fA_{\mu}$, and $G_{\mu \nu}=-ig\partial_\mu G_\nu^a \frac{\lambda_a}{2}+ig\partial_\nu G_\mu^a \frac{\lambda_a}{2}-g^2[G_\mu^a \frac{\lambda_a}{2}, G_\nu^b \frac{\lambda_b}{2}]$, and we used that $[D_\mu, D_\nu]=iq_fF_{\mu\nu}+G_{\mu \nu}$.

Once the Jacobian is regularized, it is convenient to change the basis to the free-wave eigenfunctions  $|\zeta \rangle$ $,\left(\slashed{\partial}|  \zeta \rangle=i\slashed{k}|  \zeta \rangle \right)$, to find

\begin{eqnarray}\label{Square-Opr-2}
I&=& \mathrm{tr}\sum_{n} \langle \phi_n |  x \rangle i \gamma_5 e^{-\slashed{D}^\dag(\theta)\slashed{D}(\theta)/M^2}  \langle x|  \phi_n \rangle  \qquad \qquad \qquad \qquad\ \nonumber
\\
&=& \mathrm{tr} \int \frac{d_E^4k}{(2\pi)^4}\int \frac{d_E^4k'}{(2\pi)^4}\sum_{n} \langle \phi_n |  \zeta \rangle   \langle \zeta |  x \rangle  i \gamma_5 e^{-\slashed{D}^\dag(\mu,\theta)\slashed{D}(\mu,\theta)/M^2}  \langle x|  \zeta' \rangle\langle \zeta'|  \phi_n \rangle \ \nonumber
\\
&=& \mathrm{tr}\int \frac{d_E^4k}{(2\pi)^4}\int \frac{d_E^4k'}{(2\pi)^4} \langle \zeta' |  \zeta \rangle   \langle \zeta |  x \rangle  i \gamma_5 e^{-\slashed{D}^\dag(\theta)\slashed{D}(\theta)/M^2}  \langle x|  \zeta' \rangle  \qquad  \qquad  \nonumber
\\
&=&\mathrm{tr}\int \frac{d_E^4k}{(2\pi)^4} e^{-ikx}  i \gamma_5 e^{-\slashed{D}^\dag(\theta)\slashed{D}(\theta)/M^2}  e^{ikx} \nonumber
\\ 
 &=&\mathrm{tr}\int \frac{d_E^4k}{(2\pi)^4} i \gamma_5 e^{-\slashed{D}^\dag(k,\theta)\slashed{D}(k,\theta)/M^2} , 
\end{eqnarray}
and similarly for $\widetilde{I}$

\begin{eqnarray}\label{Square-Opr-3}
\widetilde{I}&=& \mathrm{tr}\sum_{n} \langle \widetilde{\phi}_n |  x \rangle i \gamma_5 e^{-\slashed{D}(\theta)\slashed{D}^\dag(\theta)/M^2}  \langle x|  \widetilde{\phi}_n \rangle  \qquad \qquad \qquad \qquad\ \nonumber
\\
 &=&\mathrm{tr}\int \frac{d_E^4k}{(2\pi)^4} i \gamma_5 e^{-\slashed{D}(k,\theta)\slashed{D}^\dag(k,\theta)/M^2} , 
\end{eqnarray}
with $\slashed{D}^\dag(k,\theta)\slashed{D}(k,\theta)$ and $\slashed{D}(k,\theta)\slashed{D}^\dag(k,\theta)$ given respectively by (\ref{Square-Opr}) and (\ref{Square-Opr1}) with $D_\mu$ replaced by $(ik_\mu+D_\mu)$. In (\ref{Square-Opr-2}) and (\ref{Square-Opr-3}) we introduced the notation
\begin{equation}\label{Matsubara sum}
 \int \frac{d_E^4k}{(2\pi)^4}=\frac{1}{\beta} \sum_{n}\int\ \frac{d^3\vec{k}}{(2\pi)^3}, \quad k_4=\frac{(2n+1)\pi}{\beta}, \quad n=0,\pm1,\pm2,..., \quad \beta=1/T
\end{equation}

At this point, we make the variable change $k_\mu \rightarrow Mk_\mu$ in (\ref{Square-Opr-2})  and (\ref{Square-Opr-3}), use them back in (\ref{Heat-Kernel}) and (\ref{Heat-Kernel1}), and take the trace and the limit $M \to \infty$. We can readily verify that  
\begin{eqnarray}\label{Trace}
\lim_{M  \rightarrow \infty}  (I+ \widetilde{I})&=&  \lim_{M  \rightarrow \infty}\left[\mathrm{tr}\int \frac{d_E^4k}{(2\pi)^4} i \gamma_5 e^{-\slashed{D}^\dag(k,\theta)\slashed{D}(k,\theta)/M^2}+\mathrm{tr}\int \frac{d_E^4k}{(2\pi)^4} i \gamma_5 e^{-\slashed{D}(k,\theta)\slashed{D}^\dag(k,\theta)/M^2}\right]
\\
&=&2\int \frac{d_E^4k}{(2\pi)^4}e^{-k^2}\left(-\frac{q_f^2}{32}t_ri\gamma^5[\gamma_\mu, \gamma_\nu][\gamma_\alpha, \gamma_\beta]F_{\mu\nu}F_{\alpha \beta}+\frac{1}{32}t_r i\gamma^5[\gamma_\mu, \gamma_\nu][\gamma_\alpha, \gamma_\beta]G_{\mu\nu}G_{\alpha \beta}\right),\nonumber
\end{eqnarray}
where $t_r$ refers to the trace in Dirac and color matrices. 

Taking advantage of the formula \cite{Niu}
\begin{equation}\label{sum-eq}
\frac{1}{\beta} \sum_{n=-\infty}^{+\infty}\ exp \left[-k^2_4 \right ]=\frac{1}{\beta} \sum_{n=-\infty}^{+\infty}\ exp \left [-\frac{(2n+1)^2\pi^2}{\beta^2}\right]=\int_{-\infty}^{\infty}\frac{dk^0}{2\pi} exp \left[-(k^0)^2\right],
\end{equation}
we can then do the integrals in (\ref{Trace}) and use them to write

\begin{eqnarray}\label{exponent}
\mathcal{I}_R+\mathcal{\widetilde{I}}_R&=& \lim_{M \rightarrow \infty}\int d_E^4x \theta(x)(I+\widetilde{I})\nonumber
\\
&=&\frac{2}{16\pi^2}\int d_E^4x \theta(x)\left(-\frac{q_f^2}{32}t_ri\gamma^5[\gamma_\mu, \gamma_\nu][\gamma_\alpha, \gamma_\beta]F_{\mu\nu}F_{\alpha \beta}+\frac{1}{32}t_r i\gamma^5[\gamma_\mu, \gamma_\nu][\gamma_\alpha, \gamma_\beta]G_{\mu\nu}G_{\alpha \beta}\right),\nonumber
\\
&=&- i\frac{N_cq_f^2}{8\pi^2}\int d^4x \theta(x)F_{\mu\nu}\tilde{F}^{\mu \nu}- i\frac{g^2}{16\pi^2}\int d^4x \theta(x)G^a_{\mu\nu}\tilde{G}_a^{\mu \nu},
\end{eqnarray}
where in the last line of (\ref{exponent}) we went to Minkowski space and used $\mathrm{tr}\gamma^5[\gamma^\mu, \gamma^\nu][\gamma^\alpha, \gamma^\beta]=-16i\epsilon^{\mu \nu \alpha \beta}$. We also introduced the dual tensors $\tilde{F}^{\mu \nu}=\frac{1}{2} \epsilon^{\mu \nu \alpha \beta}F_{\alpha \beta}$ and  $\tilde{G}^{\mu \nu}=\frac{1}{2} \epsilon^{\mu \nu \alpha \beta}G_{\alpha \beta}$, and used that $t_rG_{\mu\nu}\tilde{G}^{\mu \nu}=-\frac{g^2}{2}G^a_{\mu\nu}\tilde{G}_a^{\mu \nu}$.

Inserting (\ref{exponent}) in (\ref{Jacobian-phi-rep_Reg}), we find that the regularized Jacobian of a single flavor is

\begin{equation}\label{Jacobian-2}
J^{(f)}_{\psi}J^{(f)}_{\bar{\psi}}=\exp \Bigg\{ i\frac{1}{N_f}\int d^4x \theta(x)\left \lbrack\frac{N_cq_f^2}{16\pi^2}F_{\mu\nu}\tilde{F}^{\mu \nu}+ \frac{g^2}{32\pi^2}G^a_{\mu\nu}\tilde{G}_a^{\mu \nu} \right\rbrack \Bigg\}
\end{equation}

Hence, taking into account the Jacobian for all the flavors, we find that the net contribution from the measure to the action is
\begin{equation}\label{Jacobian-net}
J_{\psi}J_{\bar{\psi}}=
\exp \Bigg\{ i\int d^4x \theta(x)\frac{e^2}{24\pi^2}F_{\mu\nu}\tilde{F}^{\mu \nu}+ i\int d^4x \theta(x)\frac{g^2}{32\pi^2}G^a_{\mu\nu}\tilde{G}_a^{\mu \nu} \Bigg\}
\end{equation}

Notice that the second term in the exponent eliminates the original $\theta$-vacuum term of (\ref{L_QCD_QED}), which was the goal we set for ourselves from the beginning of this section. At the same time, a new axion term has emerged, one that now mixes $\theta$ with the electromagnetic field $F_{\mu\nu}$.

After incorporating the Jacobian into the partition function (\ref{path-int}), the fermion effective action becomes,
\begin{eqnarray}\label{Action-Effective}
S_{\psi}(G_\mu^a, A_\mu, \theta)&=&\int d^4x \left[\bar{\psi}i\gamma^{\mu}(\partial_\mu-igG_\mu+iQA_{\mu}+\mu \delta_{\mu 0}+i\gamma^5\partial_\mu\theta/6)\psi+\theta(x)\frac{e^2}{24\pi^2}F_{\mu\nu}\tilde{F}^{\mu \nu}\right ]\nonumber
\\
&=&\int d^4x \lbrack\bar{\psi}i\gamma^{\mu}(\partial_\mu-igG_\mu+iQA_{\mu}+\mu \delta_{\mu 0}+i\gamma^5\partial_\mu\theta/6)\psi-\frac{e^2}{6 \pi^2} \partial_\mu \theta(x)\epsilon^{\mu \alpha \nu \beta}A_\alpha \partial_\nu A_{\beta} \rbrack\nonumber
\\
&=&\int d^4x \lbrack\bar{\psi}i\gamma^{\mu}(\partial_\mu-igG_\mu+iQA_{\mu}+\mu \delta_{\mu 0}+i\delta_{\mu0}\gamma^5\mu_5)\psi -\frac{e^2}{\pi^2} \mu_5\epsilon^{0 \alpha \nu \beta}A_\alpha \partial_\nu A_{\beta} \rbrack \nonumber
\\
\end{eqnarray}
where in the second line we integrated by part the last term for convenience, and in the third line, to touch base with the CME studies, we assumed that $\theta$ only depends on time and used $\partial_\mu \theta(x)/6=\mu_5$, for a constant $\mu_5$, in agreement with the case considered in \cite{CME}.

Notice that because of the induced Abelian axion term $\frac{\theta e^2}{24\pi^2}F_{\mu\nu}\tilde{F}^{\mu \nu}$ in the action, the electromagnetism in this theory will be described by the equations of Axion Electrodynamics.

\section{Charges and currents in hot QCD at $B\neq0$}

In this section we shall calculate all the contributions to the electric four-current. We start from the  effective action of the gauge fields

\begin{equation}\label{Gamma-total}
\Gamma(A,G)=-\frac{1}{4}\int d^4x F_{\mu \nu} F^{\mu \nu}-\frac{e^2 \mu_5}{\pi^2}\int d^4x  \epsilon^{0 \alpha \nu \beta}A_\alpha \partial_\nu A_\beta-i\ln\mathcal{Z}
\end{equation}
with 
\begin{equation}\label{PFunction-final}
\mathcal{Z}=\int \mathcal{D}\psi \mathcal{D}\bar{\psi}e^{i  \int d^4x \lbrack\bar{\psi}i\gamma^{\mu}(\partial_\mu-igG_\mu+iQA_{\mu}+\mu \delta_{\mu 0}+i\delta_{\mu0}\gamma^5\mu_5)\psi},
\end{equation}
integrate in the fermion fields and make an expansion in powers of all the gauge fields, so that $\Gamma(A,G)$ becomes

\begin{eqnarray} \label{EA-2}
\Gamma(A,G)&=&-V\Omega+\int d^4x \left[-\frac{1}{4}F_{\mu\nu}F^{\mu\nu}-\frac{e^2 \mu_5}{\pi^2}  \epsilon^{0 \alpha \nu \beta}A_\alpha \partial_\nu A_\beta\right]
\\
&+ &\sum_{n=1}^{\infty}\int dx_{1}...dx_{n}\Pi_{\mu_1,\mu_2,...\mu_n}(x_1,x_2,...x_n)A^{\mu_1}(x_1)...A^{\mu_n}(x_n)+ ... \nonumber
\end{eqnarray}
with V the four-volume, $\Omega=\Omega(T,B)$ the thermodynamic potential,  $\Pi_{\mu_1,\mu_2,...\mu_n}$ the one-loop polarization operators with internal lines of fermions and $i$ external lines of photons, and $\cdots$ indicates terms containing powers of the gluon field, which are not relevant for our goal.

We are interested in the linear response of the magnetized and hot QCD plasma to a small electromagnetic probe $\tilde{A}$. Furthermore, for consistency of the approximation, we can neglect all the radiative corrections of order higher than $\alpha$, as $\alpha$ is the order of the axion term in (\ref{Gamma-total}).  These two conditions imply that we shall cut the series in (\ref{EA-2}) at $n=1$, which can be shown to provide the medium corrections to the Maxwell equations that are linear in the electromagnetic field and of the desired order in $\alpha$.
Thus, we keep in the polarization operator series the $n=1$ order and reduce (\ref{EA-2}) to
\begin{equation} \label{EA}
\Gamma(A)\simeq-V\Omega+\int d^4x \left[-\frac{1}{4}F_{\mu\nu}F^{\mu\nu}-\frac{e^2 \mu_5}{\pi^2} \epsilon^{0 \alpha \nu \beta}A_\alpha \partial_\nu A_\beta\right]
-\int d^4x A_\mu(x) J^\mu(x),
\end{equation}
where $J^\mu(x)=(J^0,\mathbf{J})$ represents the contribution of the ordinary (non-anomalous) electric four-current, determined by the one-loop tadpole diagrams.  

\subsection{Anomalous charges and currents at $B\neq0$}

The anomalous charge and current densities are derived from the third term in the right-hand-side of (\ref{EA}). For a magnetic field in the $z$ direction ($B=-F_{12}$) this term reduces to
 \begin{equation}\label{anom-charge-term}
\Gamma_{anom}=\frac{e^2 \mu_5}{\pi^2}\int d^4xA_3B
\end{equation}

Taking the variation with respect to the electromagnetic potential we obtain
 \begin{equation}\label{Anom-Charge-2}
J_{anom}^0=0
\end{equation}
 
 \begin{equation}\label{Anom-Current-2}
 {J^3}_{anom}=-\kappa(\frac{\partial \theta}{dt}){B}=-\frac{e^2 \mu_5}{\pi^2}{B},
\end{equation} 
where the coefficient $\kappa$ of Eq. (\ref{Anom-Current}) is given by $\kappa={e^2}/{6 \pi^2}$.

The anomalous current density (\ref{Anom-Current-2}) can be seen as a polarization current associated with a  time-dependent and linear in the magnetic field electric polarization $\mathbf{P}$,
 \begin{equation}\label{Anom-Current-3}
 \mathbf{J}_{anom}= \partial_t \mathbf{P}=- \frac{e^2\mathbf{B}}{6 \pi^2}\partial_t \theta,  \quad \Rightarrow \quad \mathbf{P}=-\left ( \frac{e^2\theta}{6 \pi^2}\right )\mathbf{B}
\end{equation}

We highlight that the anomalous current runs opposite to the magnetic field. 

Eq. (\ref{Anom-Current-3}) shows that the hot QGP in a magnetic field exhibits linear magnetoelectricity. This is a consequence of the P-symmetry breaking produced by the chiral anomaly term. The magnetoelectricity here has some similarity to that found in the MDCDW phase \cite{Ferrer-Incera-PLB, Ferrer-Incera-NPB}, where P is also broken, but different from the one found in the magnetic-CFL  phase of color superconductivity \cite{MCFL}, where P is not broken and the effect is a consequence of an anisotropic electric susceptibility \cite{ME-MCFL}, so it is not linear. 

\subsection{Ordinary charge and current at $B\neq0$}

The ordinary current density $J^\mu$ with chiral chemical potential $\mu_5$ can be found with different methods.  The magnetic field is assumed to be along the $x_3$ direction. As shown in  \cite{CME}, only the LLL contributes to the it. Below, we reproduce the same result but using the tadpole diagram from the last term of (\ref{EA}). 

\textbf{Quark propagator in the LLL}

First, we need to find the LLL propagator of each quark flavor. The LLL inverse quark propagator is
\begin{equation}\label{inverse-propagator}
G^{-1}_{LLL}(p)=\gamma_\|^\mu(\tilde{p}_\mu^\|+{\mu}_5 \delta_{\mu 0}\gamma^5)-m,
\end{equation}
where $\tilde{p}_\mu^\|=(p_0-\mu, p_3)$ . It is easy to verify that the propagator $G_{LLL}(p)$ that satisfies 
\begin{equation}\label{product-propagators}
G^{-1}_{LLL}(p)G_{LLL}(p)=G_{LLL}(p)G^{-1}_{LLL}(p)=I,
\end{equation}
can be written as
\begin{equation}\label{propagator-1}
G_{LLL}(p)=\frac{AB}{\det G^{-1}_{LLL}(p)},
\end{equation}
with
\begin{eqnarray}\label{A}
A&=&-\gamma^5G^{-1}_{LLL}(p)\gamma^5\nonumber
\\
&=&\gamma_\|^\mu(\tilde{p}_\mu^\|+{\mu}_5 \delta_{\mu 0}\gamma^5)+m
\end{eqnarray}
and
\begin{equation}\label{B}
B=-\gamma^1A\gamma^1=(\tilde{p}_\|^2-{\mu}_5^2-m^2)I+2i{\mu}_5p_3(i\gamma^1\gamma^2).
\end{equation}

It is convenient to express it as a combination of the spin projectors $\Delta(\pm)=(I\pm i\gamma^1\gamma^2)/2$,
\begin{equation}\label{propagator-3}
G_{LLL}(p)=\frac{\gamma^\|_\mu \tilde{p}^\mu_++m}{(\tilde{p}_0)^2-\varepsilon^2_+}\Delta(+)+\frac{\gamma^\|_\mu \tilde{p}^\mu_-+m}{(\tilde{p}_0)^2-\varepsilon^2_-}\Delta(-),
\end{equation}
where $\tilde{p}^\nu_{\pm}=(p^0-\mu, 0, 0, p^3\pm{{\mu}}_5)$, $\gamma_\nu^\|=(\gamma_0,0,0,\gamma_3)$ and $\varepsilon_\pm=\sqrt{(p_3\pm{\mu}_5)^2+m^2}$.

Keeping in mind that the LLL quarks only have one spin projection (parallel/antiparallel to the field for positive/negative charged quarks), the LLL propagator reduces to

\begin{equation}\label{propagator}
G_{(f)LLL}(p)=G_{LLL}(p)\Delta(\mathrm{sgn}\left(q_f\right)).
\end{equation}

\textbf{The tadpole contribution}

The tadpole diagram in the LLL for each flavor contributes to the four-current as 
\begin{eqnarray}\label{4-current}
J^\mu_{(f)LLL}(\mathrm{sgn}\left(q_f\right))&=&-\frac{q_f|q_fB|N_c}{(2\pi)^2\beta}\sum_{p_4} \int_{-\infty}^{\infty} dp_3 tr \left [i\gamma^\mu G^E_{(f)LLL}(k) \right]\nonumber
\\
&=&-\frac{q_f|q_fB|N_c}{(2\pi)^2\beta}\sum_{p_4} \int_{-\infty}^{\infty} dp_3 tr\left [\gamma^\mu \frac{\gamma^4(p^4+i\mu)+\gamma^3(p^3+{\mu}_5)-m}{(p^4+i\mu)^2+\varepsilon^2_{\mathrm{sgn}\left(q_f\right)}}\Delta(\mathrm{sgn}\left(q_f\right)) \right],
\end{eqnarray}
where we did the Wick rotation to Euclidean space and introduced the Matsubara's sum with $p_4=\frac{(2n+1)\pi}{\beta}, n=0,1,2,...$, $\beta=1/T$.

Taking the trace in (\ref{4-current}), we have 

\begin{equation}\label{trace-0}
tr\left[\gamma^\nu(\gamma_\|^\mu \tilde{p}^\mu_\pm-m)\Delta(\pm)\right ]=-2\tilde{p}^4, \qquad \nu=4
\end{equation}

\begin{equation}\label{trace-1}
tr\left[\gamma^{\nu}(\gamma_\|^\mu \tilde{p}^\mu_\pm-m)\Delta(\pm)\right ]=0, \qquad \quad \nu=1,2
\end{equation}

\begin{equation}\label{trace-2}
tr\left[\gamma^\nu(\gamma_\|^\mu \tilde{p}^\mu_\pm-m)\Delta(\pm)\right ]=-2\tilde{p}^3, \qquad \nu=3
\end{equation}
with $\tilde{p}^\nu_{\pm}=(p^4+i\mu, 0, 0, p^3\pm{\mu}_5)$ in Euclidean space.

We find that the LLL does not contribute to the ordinary transverse electric current density ($J_{LLL}^{1,2}=0$) due to the zero trace (\ref{trace-1}). Hence, only a current along the magnetic-field direction and a charge density can have nontrivial values. 

\textbf{Ordinary charge density}

The LLL contribution to the ordinary electric charge of each quark flavor is  obtained substituting  (\ref{trace-0}) in (\ref{4-current}) as
\begin{equation}\label{4-current-LLL}
J^4_{(f)LLL}(\mathrm{sgn}\left(q_f\right))=\frac{q_f|q_fB|N_c}{2\pi^2\beta}\sum_{p_4}  \int_{-\infty}^{\infty} dp_3 \frac{2({p}^4+i\mu)}{({p}^4+i\mu)^2+\varepsilon^2_{sgn(q_f)}}, 
\end{equation}
with $\varepsilon^2_{sgn(q_f)}=(p_3+sgn(q_f){\mu}_5)^2+m^2$.
Carrying out the Matsubara sum in (\ref{4-current-LLL}), and making the analytic continuation to Minkowski space,  we obtain
\begin{equation}\label{4-current-T}
J^0_{(f)LLL}(\mathrm{sgn}\left(q_f\right))=\frac{q_f|q_fB|N_c}{2\pi^2} \int_{-\infty}^{\infty} dp_3 \left [  n_F\left[\beta(\varepsilon_{sgn(q_f)}+\mu)\right]
-n_F\left[\beta(\varepsilon_{sgn(q_f)}-\mu)\right] \right],
\end{equation}
where $n_F(\beta x)=[1+\exp(\beta x)]^{-1}$ is the Fermi-Dirac distribution.

Notice that the electric charge does not depend on $\mu_5$, as can be easily seen by a variable change in $p_3$ in (\ref{4-current-T}). From (\ref{4-current-T}), it is evident that if the baryon chemical potential $\mu$ is zero, the electric charge density is zero. It is then consistent to neglect the electric charge in the hot QGP system, where the corrections $(\mu/T)^n$ are really very small. 

\textbf{Ordinary current density}

The LLL contribution of a single flavor to the 3-component of the ordinary electric current is obtained substituting (\ref{trace-2}) in (\ref{4-current}), 
\begin{equation}\label{current-LLL}
J^3_{(f)LLL}(\mathrm{sgn}\left(q_f\right))=\frac{q_f|q_fB|N_c}{2\pi^2\beta}\sum_{p_4}  \int_{-\infty}^{\infty} dp_3 \frac{p^3+\mathrm{sgn}\left(q_f\right){\mu}_5}{({p}_4+i\mu)^2+\varepsilon^2_{sgn(q_f)}}, 
\end{equation}
with $\varepsilon^2_{sgn(q_f)}=(p^3+sgn(q_f){\mu}_5)^2+m^2$.
Carrying out the Matsubara sum in (\ref{current-LLL}), and returning to Minkowski space,  we obtain
\begin{eqnarray}\label{current-T}
J^3_{(f)LLL}(\mathrm{sgn}\left(q_f\right))&=&\frac{q_f|q_fB|N_c}{(2\pi)^2} \int_{-\Lambda}^{\Lambda} dp^3 \frac{p^3+sgn(q_f){\mu}_5}{\varepsilon_{sgn(q_f)}}
\left [ 1- \sum_\pm n_F\left[\beta(\varepsilon_{sgn(q_f)}\pm\mu)\right]\right]\nonumber
\\
&=&\frac{q_f|q_fB|N_c}{(2\pi)^2} \int_{-\Lambda}^{\Lambda} dp^3\frac{d}{dp^3}\left[\varepsilon_{sgn(q_f)}+\sum_\pm\frac{1}{\beta}\log n^{-1}_F\left[-\beta(\varepsilon_{sgn(q_f)}\pm\mu)\right]
\right ]\nonumber
\\
&=&\frac{q_f|q_fB|N_c}{2\pi^2}sgn(q_f) {\mu}_5
\end{eqnarray}

Summing in flavor, we find the net ordinary current to be
\begin{equation}\label{regular-current}
J_{LLL}^3(\mu_5)=\sum_{f=u,d,s} J^3_{LLL}(sgn(q_f))=\frac{e^2{ \mu}_5}{\pi^2}B
\end{equation}
If we take $q_f=e$ and $N_c=1$ in (\ref{current-T}), we find the same expression obtained in \cite{CME} for the current of a single fermion with charge e and "chiral chemical potential" $\mu_5$. Notice, that the ordinary current $J_{LLL}^3$ in (\ref{regular-current}) is equal to \emph{minus} the anomalous current density (\ref{Anom-Current-2}), so they cancel out in (\ref{axQED_2}), as we had previously announced.  Itis worth to underline that the particle mass, temperature and baryonic chemical potential do not make any contribution to this current. 

\section{CME and AHE in Weyl semimetals}

Weyl semimetals are three-dimensional, topologically nontrivial materials that present linearly dispersing quasiparticle excitations of opposite chiralities in the vicinity of the two so-called Weyl points, which are nodes separated in momentum space \cite{WS}.

Around a Weyl node, the low-energy theory of the WSM in Euclidean space is described by the action \cite{Zyuzin-Burkov} 
\begin{equation}\label{WS-action}
S= \int d^4x_E \bar{\psi}\left[i\gamma^\mu(\partial_\mu+ieA_\mu+ib_\mu\gamma^5)\right ]\psi
\end{equation}
with $b_\mu=(\textbf{b}, b_4)$, $\textbf{b}$ being the separation in momentum of the two Weyl nodes, and $b_4$ the separation in energy. 

The action (\ref{WS-action}) describes a gapless system, but similarly to QCD, under a strong coupling regime, the system can become gapped \cite{gaped}. Nevertheless, the  coupling critical value needed to produce the condensate that makes the theory gapped is too high to be realized in condensed matter systems. On the other hand, in the presence of a magnetic field the condensate can be induced even in the weak-coupling regime, similar to the phenomenon of magnetic catalysis of chiral symmetry breaking that can take place in QED and QCD \cite{MC}. In this case, however, due to the momentum-space separation of the Weyl nodes, the chiral symmetry breaking occurs through a momentum-dependent particle-hole condensate that breaks translational invariance and represents a charge-density-wave \cite{B-condensate}. Nevertheless, the dynamical gap generated by magnetic catalysis is always a very small parameter \cite{MC} and can be neglected, thus the action (\ref{WS-action}) is a good approximation. 

Now, the $b_\mu$ term can be eliminated in (\ref{WS-action}) by the gauge chiral transformation
\begin{eqnarray}\label{chiral-transformation}
\psi(x)= e^{-i\theta(x) \gamma^5/2}\psi(x)\nonumber
\\
\bar{\psi}(x) =\bar{\psi}(x)e^{-i\theta(x) \gamma^5/2}
\end{eqnarray}
with $\theta(x)=2b_\mu x_\mu$. But as in the QCD case, the partition function measure is not invariant and the Jacobian, after regularization by the Fujikawa's method, gives rise to a contribution $\theta(x) F_{\mu \nu}\tilde{F}^{\mu \nu}$ to the action \cite{Zyuzin-Burkov}, 
\begin{equation}\label{WS-action-2}
S= \int d^4x_E \left\{\bar{\psi}\left[i\gamma^\mu(\partial_\mu+ieA_\mu)\right ]\psi+\frac{ie^2}{32 \pi^2} \theta(x)F_{\mu \nu}\tilde{F}^{\mu \nu} \right \}
\end{equation}

The $\theta$-vacuum term in (\ref{WS-action-2}) produces all the anomalous contributions for charge and currents we discussed in Section II. We note that the covariant derivative in (\ref{WS-action-2}) does not depend on $\theta$, so the ordinary charge and currents cannot eliminate here the anomalous contributions coming from the $\theta$-vacuum term.
We conclude that WSM are natural systems where in principle both anomalous transport phenomena: CME and AHE, could be realized. 

Notwithstanding, the realization of the CME in WSM has been also contested. In \cite{Franz} it was argued that the CME is an artifact of linearizing the quasiparticle dispersion relations near the nodes, but that going beyond the low-energy approximation in the full lattice model, it is absent. On the other hand, the AHE was found to remain robust in realistic Weyl semimetals defined on the lattice \cite{Franz}. In another direction, it was shown in \cite{Limits}, that the existence or not of the CME depends on the order the zero limits of frequency and momentum are taken: $\omega \rightarrow 0$ and $\textbf{k} \rightarrow 0$. If the static limit $\omega \rightarrow 0$ is taken first,  the system is in equilibrium and the CME is absent as predicted in \cite{Franz}. On the contrary, by taking first the plasmon limit, $\textbf{k} \rightarrow 0$, the CME survives. In this case the system is in non-equilibrium. There was also an extra physical argument \cite{Basar}, based on energy reasonings, to discard the CME in the equilibrium system. 

\section{Concluding remarks}
 
We have shown that in the presence of a magnetic field, the QCD  $\theta$-vacuum with a time-dependent $\theta=2N_f\mu_5t$, gives rise to a couple of currents (anomalous and ordinary) moving in opposite directions along the magnetic field that cancel out each other. The anomalous current (\ref{Anom-Current-2}) is produced by the time-dependent medium polarization induced by the axion term of the action. This current is opposite to the field direction. On the other hand, the ordinary current (\ref{regular-current}) is produced by the motion of the quarks seating in the LLL that in this case gives rise to a net motion of positively charged particles in the  field direction and of negative charges in the opposite direction. This mechanism \cite{Warringa} has been considered as the hallmark of the CME effect and it relies on the chirality imbalance generated in the hot QGP and the single spin of the LLL quarks in an applied magnetic field. 

Any of the two currents alone (anomalous or ordinary), by considering a time-depending polarization or a stable drift of quarks with different chiralities, would imply a situation of non-equilibrium emerging within a treatment that has assumed thermodynamic equilibrium, what would be contradictory. The fact that the two currents are equal and opposite, ensures an equilibrium situation. Therefore, the electromagnetism in this case    

It is worth to mention that in several recent publications the existence of the CEM is equilibrium was also questioned. In \cite{Lattice}, using lattice field theory, the CME current was calculated and found that in the equilibrium bulk, the CME does not exist, while the CME current may appear close to the boundary in the case of a finite-size system, although the integrated total CME current remains zero. In a different direction, in \cite{Zubkov}, on the basis of the Wigner 's transform technique applied to Green functions, it was found that the equilibrium CME current is also absent in the properly regularized quantum field theory.

We argued that the lack of CME in the hot QGP in equilibrium is a consequence of the trivial topology of the fermions in this system. The topology in this case is associated with the gluon vacuum, and it never gets transferred to the fermions, whose spectrum remains symmetric, so topologically trivial, even after the local chiral transformation. That is why the topology of the gluon vacuum cannot leave an imprint into the electromagnetism of the system. This explains why even considering the baryon chemical potential, the electric charge is independent from the chiral chemical potential, and why no macroscopic effect like the CME current exists in equilibrium. This is in sharp contrast with the situation in the MDCDW phase of QCD at finite density. There, the topology of the system comes from the fermion groundstate and affect the fermion spectrum, which is asymmetric for the LLL quarks. As a consequence, the anomalous Hall current is not cancelled by the ordinary one.  The anomalous current and hence the anomalous transport in the MDCDW phase is robust because it has a topological origin \cite{Ferrer-Incera-PLB}-\cite{Ferrer-Incera-NPB}. We speculate that the same analysis applies to the WSMs, where, as discussed in the last section, the AHE remains robust even when one goes beyond the low-energy approximation, while the CME does not.

\end{document}